\documentclass[aps,prb,twocolumn,superscriptaddress]{revtex4}

\usepackage[dvips]{graphicx}
\begin{document}


\title{Signatures of pressure induced superconductivity in insulating Bi2212}


\author{T. Cuk}
\affiliation{Department of Applied Physics, Stanford University,
Stanford, CA 94305}
\author{D. A. Zocco}
\affiliation{Department of Physics, University of California, San Diego,
La Jolla, California 92093, USA}
\author{H. Eisaki}
\affiliation{Nanoelectronics Research Institute, National
Institute of Advanced Industrial Science and Technology, 1-1-1
Central 2, Umezono, Tsukuba, Ibaraki, 305-8568, Japan}
\author{V. Struzhkin}
\affiliation{Carnegie Institution of Washington: Geophysical
Laboratory 5241 Broad Branch Rd NW Washington, DC 20015 }
\author{M. Grosche}
\affiliation{Department of Physics, Royal Holloway, University of
London, Egham, Surrey TW20 0EX}
\author{M. B. Maple}
\affiliation{Department of Physics, University of California, San Diego,
La Jolla, California 92093, USA}
\author{Z.-X. Shen}
\affiliation{Departments of Physics, Applied Physics, and Stanford
Synchrotron Radiation Laboratory, Stanford University, Stanford,
CA 94305}


\date{February 19, 2010}

\begin{abstract} We have performed several high pressure electrical resistance experiments on Bi$_{1.98}$Sr$_{2.06}$Y$_{0.68}$Cu$_{2}$O$_{8+\delta}$, an insulating parent compound of the high-T$_{c}$ Bi2212 family of copper oxide superconductors.  We find a resistive anomaly, a downturn at low temperature, that onsets with applied pressure in the 20-40 kbar range. Through both resistance and magnetoresistance measurements, we identify this anomaly as a signature of induced superconductivity.  Resistance to higher pressures decreases T$_{c}$, giving a maximum of $\sim$ 10 K. The higher pressure measurements exhibit a strong sensitivity to the hydrostaticity of the pressure environment.  We make comparisons to the pressure induced superconductivity now ubiquitous in the iron arsenides.
\end{abstract}

\pacs{}

\maketitle 
\section{Introduction}
The recently discovered class of iron arsenide superconductors have been compared to the high-T$_{c}$ copper-oxide superconductors based on the proximity of the superconducting dome to an antiferromagnetic phase in the doping phase diagram and the planar-like sheets in which the superconductivity lies.\cite{Rotter}  In both classes of high-T$_{c}$ families, the phase diagram exhibits a large range of structural, charge, and magnetic instabilities that manifest themselves more strongly in one specific compound or at a particular doping than another.  Given the complexities of understanding how physical parameters are changed by chemical doping, pressure has been sought as a laboratory controlled variable to tune a single parent compound accross the phase diagram.  In the iron arsenide compounds, pressure induced superconductivity has already been found in CaFe$_{2}$As$_{2}$,\cite{CaFeAs} BaFe$_{2}$As$_{2}$,\cite{Alireza} SrFe$_{2}$As$_{2}$,\cite{Alireza} and LaFeAsO.\cite{LaFeAsO} The pressure induced superconductivity appears as the tetragonal to orthorhombic structural transition temperature is lowered and the associated spin and charge density wave transitions weaken.\cite{SrFeAs} In the cuprates, by contrast, pressure induced superconductivity of the parent, or weakly doped, compounds has not been found.

Similar to the iron arsenide compounds, there are several mechanisms by which
pressure could induce metallicity in the cuprate parent insulators. 
By decreasing the Cu-O bond distance, the
Cu$_{d}$-O$_{p}$ hybridization increases and the Cu$_{d}$-band
further delocalizes. Secondly, the relatively higher \textsl{c}-axis compressibility facilitates electron or hole doping by bringing the chemically substituted charge resevoir layers
closer to the Cu-O plane.\cite{Murayama} Preferred compression along the layered
\textsl{c}-axis also increases the interlayer tunneling matrix element
(t$\bot$).\cite{Forro, Crommie}  Finally, contraction of the lattice
changes electron-electron and electron-phonon interactions by
varying both electron screening and crystal fields.\cite{Chen}

Past pressure experiments on superconducting
Bi$_{2}$Sr$_{2}$CaCu$_{2}$O$_{8}$ have indeed shown significant
increase of T$_{c}$ on the underdoped side of the superconducting dome,\cite{Chen} attributed to a pressure-induced doping as inferred from Hall
effect measurements.\cite{Murayama} However, pressure leads to
higher T$_{c}$'s for even overdoped Bi2212 compounds, which suggests that the strength of the Cooper-pair coupling also changes.\cite{Chen} Similar results are found for
YBa$_{2}$Cu$_{3}$O$_{7}$.\cite{Jorgensen, Schilling, Murayama,
Tozer} The sensitivity of the \textsl{c}-axis resistivity in
Bi$_{2}$Sr$_{2}$(Ca,Y)Cu$_{2}$O$_{8}$ and YBa$_{2}$Cu$_{3}$O$_{7}$
to uniaxial pressure has been attributed to increased interlayer
tunneling t$\bot$.\cite{Forro, Crommie}

Recently, pressure has tuned insulating Bi2212 accross an electronic transition
seen by Raman spectroscopy at high (200 kbar) pressures.\cite{Cuk} The electronic transition
coincides with a change in the behavior of the magnons, the phonons, and the \textsl{c/a}
lattice constant ratio.  So far, however, pressure induced superconductivity has not
been reported in a parent cuprate compound.  In this work, we study the effect of pressure on insulating Bi2212 in
resistance measurements using two different Bridgman anvil cell setups and a diamond anvil cell.  We identify a broad downturn at $\sim$ 8-10 K occuring at $\sim$30 kbar, as a signature of induced superconductivity. In all three experiments, the resistive downturn indicative of superconductivity is observed
to have an onset in the 20-40 kbar range.  However, the high pressure range (above 40 kbar) differs due to the
variations in hydrostaticity.  A higher \textsl{c}-axis stress is correlated with a more extended range of 
induced superconductivity and an enhanced metallic behavior.  Finally, we compare these results to the pressure induced superconductivity seen in the iron arsenides.  

\section{Experimental Details}
Single crystals of
Bi$_{1.98}$Sr$_{2.06}$Y$_{0.68}$Cu$_{2}$O$_{8+\delta}$ (Bi2212)
were grown by the floating zone method, and have a doping
dependence of T$_{c}$ described by Maeda \cite{Maeda} and Terasaki.\cite{Terasaki} Given the sensitivity of these samples to 
the hydrostaticity of the pressure environment, we report results using three
different pressure configurations.  
In the first, a sample of 1$\times$0.25$\times$0.025 mm$^{3}$ was loaded in a beryllium-copper Bridgman-anvil clamped cell (BAC1) using solid steatite as the quasihydrostatic pressure transmitting medium. The pressure was determined from the superconducting transition of a strip of Pb foil placed adjacent to the sample. Electrical resistivity was measured with a standard four point technique and a LR-700 AC resistance bridge, using four flattened 50 micron platinum wires for each the sample and the Pb. Silver pads of DuPont 7095 silver paste were made in order to achieve better electrical contacts. This was the most hydrostatic configuration achieved.  In the second configuration,
a sample of 0.8$\times$0.2$\times$0.02 mm$^{3}$
(LWH) was also loaded into a Bridgman anvil cell (BAC2) next to a Pb manometer. Calcium sulphate powder served as a transparent pressure medium instead. The electrical contacts for the sample and Pb manometer were obtained as in the BAC1 experiment. The magnetoresistance measurements reported for BAC2 were preformed with a Quantum Design Physical Property Measurement System (PPMS) instrument. In BAC2, we report results from the two point measurements alone due to
complications with a non-homogeneous pressure environment. The width of the Pb transition is a measure of pressure inhomogeneity along the Pb manometer: $\pm$ 2 kbar in BAC1 and $\pm$ 5 kbar in BAC2. In the third configuration, the samples
were loaded into a Diamond Anvil Cell (DAC) with argon gas as the pressure transmitting medium. 
The $\sim$ 0.03$\times$0.03$\times$0.005 mm$^{3}$
Bi2212 samples were glued onto one of the diamond anvils and Pt
leads were, with a Focus Ion Beam, milled directly onto the top
and side surfaces of the sample in a four-point configuration and then extended along the
diamond to meet the gasket leads. The gasket
was made of a boron nitride/epoxy mixture and ruby chips served as pressure markers. In
the DAC, measurements of ruby chips placed at opposite ends of the
sample estimate the degree of non-hydrostaticity/ \textsl{c}-axis uniaxial pressure to be 20 kbar at the highest pressures. In all experimental configurations, pressure was increased at
room temperature; pressure values were recorded at low temperature since the pressure remains almost constant below 100K. 

From BAC1 to BAC2 to the DAC experiments, the degree of non-hydrostaticity/ \textsl{c}-axis uniaxial stress, increases. The contact configuation and pressure medium determine the level of hydrostaticity.\cite{Lee} In BAC1, steatite and thin Pt wires are used as opposed to the less hydrostatic calcium sulphate and thicker Pt wires in BAC2.  In the DAC experiments, while argon can be more hydrostatic than a solid powder medium in the lower pressure ranges, it solidifies at the measured temperatures and pressures (below 80 K at ambient pressure). \cite{revsciinst, jappcryst}  Moreover, in the DAC experiments, the sample is glued to the diamond surface on which the Pt leads are milled; only one side of the sample sees the quasi-hydrostatic medium.  The pressure gradients measured via Ruby fluorescence measurements in the DAC (20 kbar), and the width of the Pb superconduting transtion in BAC1 (2 kbar) and BAC2 (5 kbar) corroborate these assessments.

\section{Results}

Figure 1 shows the temperature dependent resistance of the sample loaded in BAC1
at different pressures.  By 15 kbar, the insulating characteristic
of the unpressurized sample has become a downturn below 8 K of 50 $\Omega$. With increasing pressure, the T$_{c}$ of the downturn lowers and the magnitude decreases (see inset). By 36 kbar, the T$_{c}$ decreases to 2 K, and by 40 kbar, the sample is completely insulating to 1.8 K.  Above 8 K, the insulating rise initially decreases with pressure, indicating a more metallic sample, but increases significantly beyond $\sim$ 30 kbar along with the overall resistance.

\begin{figure}
	\includegraphics[width=0.50\textwidth]{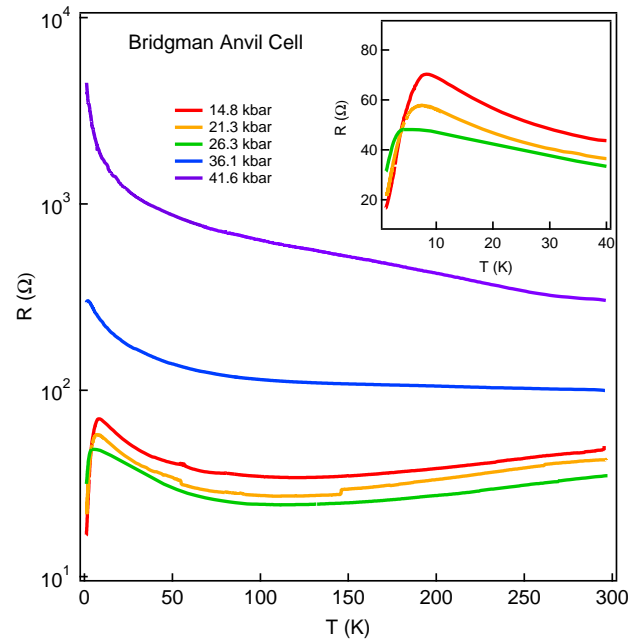}
	\caption{Resistance vs temperature shown on a log scale for different pressures for 				
	         Bi$_{1.98}$Sr$_{2.06}$Y$_{0.68}$Cu$_{2}$O$_{8+\delta}$ 
					 loaded in BAC1.  Inset: Low temperature resistance drop on a linear scale for selected pressures.}
	\label{fig:Figure1}
\end{figure}

\begin{figure}
	\centering
		\includegraphics[width=0.50\textwidth]{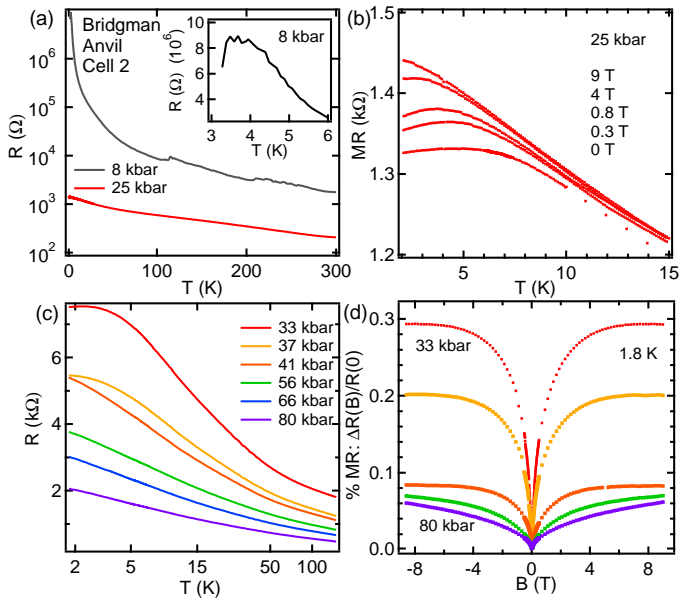}
	\caption{Bi$_{1.98}$Sr$_{2.06}$Y$_{0.68}$Cu$_{2}$O$_{8+\delta}$ loaded in BAC2 (a) Resistance vs temperature for two selected pressures in the lower pressure range
	(b) Magnetoresistance vs temperature for several different magnetic fields at 25kbar, where the low temperature resistive
	anomaly was observed (c) Resistance vs temperature for the higher pressure range plotted on a ln T scale (d) Magnetic
	field sweeps for the curves shown in (c).}
	\label{fig:Figure2}
\end{figure}

Figure 2 shows the temperature dependent resistance and magnetoresistance of the sample loaded in BAC2
at different pressures. In Fig. 2a, the resistance decreases by four orders of magnitude between 8 kbar and
25 kbar. Though not as pronounced as in BAC1, the insulating characteristic becomes a downturn at low temperatures---most pronounced at 25 kbar (Fig. 2b) but also apparent at 8 kbar (see inset to Fig. 2a).
In these experiments, we applied a magnetic field and found that the resistive anomaly was almost totally suppressed with an applied field of 9 T (Fig. 2b).  We use this positive magnetoresistance (increasing resistance with applied magnetic field) as another measure of the induced superconductivity.  Since the magnon peak is not suppressed in this pressure range, as seen earlier by pressure dependent Raman studies,\cite{Cuk} the positive magnetoresistance cannot be attributed to a suppression of short range antiferromagnetic order.\cite{LongAF} We show sweeps in magnetic field for several pressures at 1.8 K in Fig. 2d. The largest magnetoresistance is observed in the range of 30 kbar and decreases significantly beyond 40 kbar. Like in BAC1, then, the induced signatures of superconductivity disappear with continued increase in pressure. The high magnetoresistance occurs for those pressures that show either a flattening or a downturn of the resistance at low temperatures (Fig. 2c).  Further increase in pressure continues to make the sample more metallic (Fig. 2c) and in contrast to BAC1, continues to become increasingly more metallic up to 80 kbar.  Beyond $\sim$ 50 kbar, we find a - ln T(K) dependence over two decades in temperature.  This same characteristic is prominent in Bi2201 samples near the dome \cite{Ong1} and when superconductivity is suppressed by the application of a magnetic field in Bi2201.\cite{Ono} 

\begin{figure}
	\centering
		\includegraphics[width=0.4\textwidth]{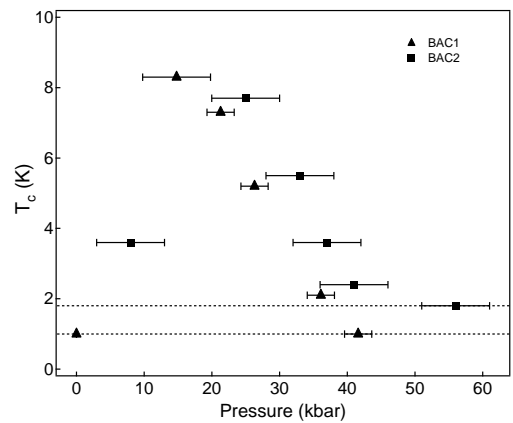}
	\caption{T$_{c}$ vs pressure for Bi$_{1.98}$Sr$_{2.06}$Y$_{0.68}$Cu$_{2}$O$_{8+\delta}$ loaded in BAC1 and BAC2.  The lowest measured temperatures for BAC1 and BAC2 are indicated by the dotted lines. Error bars on the pressure points are taken from the width of the Pb superconducting transition and indicate the uniaxial stress along the c-axis.}
	\label{fig:Figure3}
\end{figure}
 
In both BAC1 and BAC2, the strongest signatures of superconductivity occur below 40 kbar.  
In Fig. 3a, we graph T$_{c}$ vs pressure for both cells. T$_{c}$ is determined by the temperature
at which the high temperature slope diverges from the low temperature behavior.  Even at 8 kbar
we find a downturn at 4 K, which suggests that as with the pressure induced superconductivity in the iron pnictides,
T$_{c}$ traces a dome with pressure.\cite{Lee, Takahashi, Diego2, Hamlin}  In the region beyond 40 kbar, the results in BAC1 and BAC2 differ substantially: in BAC1, we observe the return of an insulating state, whereas in BAC2, we find resistance curves similar to those of doped samples near the dome.

\begin{figure}
	\centering
		\includegraphics[width=0.4\textwidth]{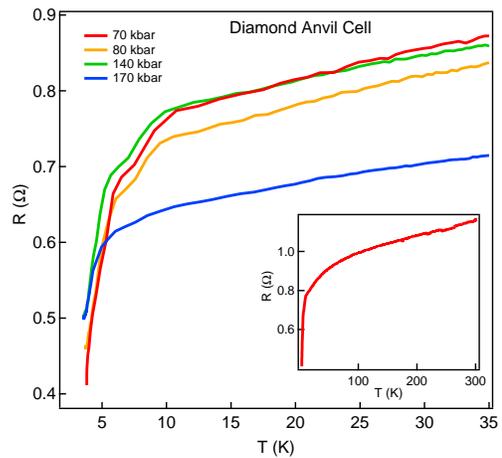}
	\caption{Resistance vs temperature for Bi$_{1.98}$Sr$_{2.06}$Y$_{0.68}$Cu$_{2}$O$_{8+\delta}$ loaded in the DAC in the low temperature range.
	Inset: Resistance curve at 70kbar for the entire temperature range.  }
	\label{fig:Figure4}
\end{figure}

Finally, in Figure 4, we show the results from the DAC experiments.  In the DAC we were unable to reach pressures lower than 70 kbar without
releasing the gas medium.  From 70 to 140 kbar, we find a downturn again indicating induced superconductivity.  With pressure, the downturn moves from 10 K to lower temperatures, reaching 5 K by 170 kbar.  Similar to the BAC results, beyond a certain pressure, the superconducting signature diminishes. However, in strong contrast to the BAC results, the superconducting signature extends to 170 kbar where in the BAC it ends at 50 kbar. In further contrast to the BAC results, above 10 K, the sample is purely metallic up to 300 K (see inset) and through the highest measured pressures.  We attribute this metallic behaviour and extended superconductivity to the higher \textsl{c}-axis stress in the DAC: greater than in the BAC due to both the solid Argon pressure medium at low temperatures and the milled Pt wires (see experimental section).  As discussed in the introduction, a higher \textsl{c}-axis stress can lead to enhanced metallic behaviour by bringing the charge resevoir layers closer to the Cu-O plane, enhancing interlayer tunneling, and changing electron-electron and electron-phonon interactions through the crystal field along the \textsl{c}-axis.  

\section{Discussion and Summary}

We find a signature of superconductivity occuring in insulating Bi2212 between 20-40 kbar in the BAC
experiments, extending to 170 kbar in the DAC.  If one considers the parent copper oxide insulators as
Mott insulators, such a low pressure onset of superconductivity is not expected.
A classic Mott insulator such as MnO only reaches metallization at 100 GPa.\cite{MnO} Therefore, these superconducting signatures are not likely the result of a Mott gap collapsing through simple hybridization of orbitals.  

Even though
superconductivity seems to onset at much lower pressures, T$_{c}$ reaches at most 10 K.  In the iron arsenide compounds, the T$_{c}$ under pressure has been shown to reach 30 K,\cite{Alireza, LaFeAsO, SrFeAs2} approximately the T$_{c}$ of the optimally doped parent compound at ambient pressure. In our experiments with Bi2212, T$_{c}$ under pressure only reaches a fraction of the optimally doped value of $\sim$ 100 K.  On the other hand, it is somewhat surprising that in both classes of high T$_{c}$ superconductors, signatures of superconductivity can be seen in the form of broad resistance curves \cite{SrFeAs2, LaFeAsO} by the same pressure of $\sim$ 30 kbar. 
One possible outlook is that the superconductivity taking place in the copper oxides with doping are of two forms: one that can be traced to lower pressures, and bearing similarity to the pressure induced superconductivity of the iron arsenides, and another higher T$_{c}$ superconductivity that has only been accessed by doping so far.  In the recent Raman spectroscopy experiments carried out in Bi2212 samples under pressure, it was possible to identify an electronic transition at 200 kbar with an anomaly occurring in the lattice constant ratio \textsl{c/a}.\cite{Cuk}  It is worth mentioning that a similar, though more dramatic, decrease occurs in \textsl{c/a} when CaFe$_{2}$As$_{2}$ enters the collapsed tetragonal state \cite{CaFeAs2} at much lower pressures.  

It is unlikely that a more hydrostatic enviroment will induce a higher T$_{c}$ in the cuprates. For the largest downturn
and highest T$_{c}$ measured in BAC1, the pressure gradients at 14 kbar approached $\pm$ 5 kbar (see Fig. 3) and only decreased at higher pressures
where the superconductivity weakened.  Moreover, we attempted the same experiments in hydrostatic piston cylinder cells and little change is observed up to 24 kbar. Such an observation is not unexpected: clear differences
are observed for CePd$_{2}$Si${_2}$ in which anvil cells report strong enhancement of T$_{c}$ and a doubled pressure range in which the superconductivity occurs compared to piston-cylinder cells. \cite{Demuer} Furthermore, comparisons with the iron arsenides also suggest that  a more hydrostatic environment may not enhance T$_{c}$.  CaFe$_{2}$As$_{2}$ does not become superconducting under ideal hydrostatic pressure conditions \cite{Lee} and internal strain coming from \textsl{c}-axis orientated planar defects in SrFe$_{2}$As$_{2}$ promotes superconductivity. \cite{Saha}

Instead, one option for future experiments is to apply a purely \textsl{c}-axis
stress to see if a more robust superconductivity can be induced.  
In particular, since the \textsl{c}-axis has the highest
compressibility and also has the highest applied stress in the
current experimental setup, one can think of the distance
between the Cu-O planes and the intervening charge resevoir layers as tuning
four separate microscopic quantities: charge transfer and disorder
(by bringing the dopants closer to the plane), interlayer
tunnelling (decreasing t$_{\bot}$), and increasing the crystal
field across the Cu-O plane.  Some of these effects, such as charge transfer and disorder or
the crystal field and charge transfer, will compete to determine the induced superconductivity.
It is an open question whether or not these competing effects can be controlled with a parameter other
than doping to achieve the optimal T$_{c}$ of the copper oxide family.


\begin{thebibliography}{04}

\bibitem{Rotter}
M. Rotter, M. Tegel, and D. Johrendt {Phys. Rev. Lett.} {\bf 101}, 107006 (2008)

\bibitem{CaFeAs} 
	 M.S. Torikachvili, S.L. Bud'ko, N. Ni, and P. C. Canfield {Phys. Rev. Lett.} {\bf 101}, 057006 (2008) 
	 	
\bibitem{Alireza}
	P. L. Alireza {\it et al.} {J. Phys.: Condens. Matter} {\bf 21}, 012208 (2009)

\bibitem{LaFeAsO}
		H. Okada {\it et al.} {J. Phys. Soc. Japan}, {\bf 77}, 113712 (2008)

\bibitem{SrFeAs}
		M.S. Torikachvili, S.L. Bud'ko, N. Ni, and P. C. Canfield {Phys. Rev. B} {\bf 78}, 104527 (2008) 
		
\bibitem{Forro}
   L. Forro, V. Ilakovac, and B. Keszei {Phys. Rev. B} {\bf 41}, {9551} (1990)

\bibitem{Crommie}
   M. F. Crommie {\it et al.} {Phys. Rev. B} {\bf 39}, {4231}(1989)

\bibitem{Chen}
   Xiao-Jia Chen, V. Struzhkin, R. J. Hemley, H. K. Mao, and C. Kendziora {Phys. Rev. B} {\bf 70}, {214502} (2004)

\bibitem{Murayama}
   C. Murayama {\it et al.} {Physica C} {\bf 183}, {277} (1991)

\bibitem{Jorgensen}
   J.D. Jorgensen {\it et al.} {Physica C} {\bf 171}, {93} (1990)

\bibitem{Schilling}
   S. Klotz, Reith, and J.S. Schilling {Physica C} {\bf 172}, {423} (1991)

\bibitem{Tozer}
   S.W. Tozer, J.L. Koston, and E. M. McCarron, III {Phys. Rev. B} {\bf 47}, {8089} (1993)
   
\bibitem{Cuk}
	 T. Cuk {\it et al.} {Phys. Rev. Lett.} {\bf 100}, {217003}(2008)

\bibitem{Maeda}
   A. Maeda {\it et al.} {Phys. Rev. B} {\bf 41}, {6418} (1990)
   
\bibitem{Terasaki}
    I. Terasaki {\it et al.} {Proceedings of the 12th Conference on Superconductivity cond-mat9911153v1 (1999)}

\bibitem{Lee}
	H. Lee {\it et al.}, {Phys. Rev. B} {\bf 80}, {024519} (2009)

\bibitem{revsciinst}
		N. Tateiwa {\it et al.}, {Rev. Sci. Inst.} {\bf 80}, {123901} (2009)

\bibitem{jappcryst}
		R. J. Angel {\it et al.}, {J. Appl. Cryst.} {\bf 40}, {26} (2007)

\bibitem{LongAF}
		Long range anti-ferromagnetic order has not yet been established in Bi2212 through neutron scattering.  

\bibitem{Ong1}
   T.W. Jing, N.P. Ong, R.V. Ramakrishnan, J.M. Tarascon, and K. Remschnig {Phys. Rev. Lett} {\bf 67}, {761} (1991)
   
\bibitem{Ono}
   S. Ono {\it et al.} {Phys. Rev. Lett.} {\bf 85}, {638} (2000)

\bibitem{Takahashi}  
	H. Takahashi {\it et. al} {Nature} {\bf 453}, {376} 
	(2008)

\bibitem{Diego2}
	D. A. Zocco {\it et al.} {Physica C} {\bf 468}, {2229} (2008)
	
\bibitem{Hamlin}
  J. J. Hamlin {\it et al.} {J. Phys: Condens. Matter} {\bf 20}, {365220} (2008)
  
\bibitem{MnO}
	 C.S. Yoo {\it et al.} {Phys. Rev. Lett.} {\bf 94}, {115502} (2005)

\bibitem{SrFeAs2}
		M. Kumar {\it et al.} {Phys. Rev. B} {\bf 78}, {184516} (2008)
		
\bibitem{CaFeAs2} 
		A.I. Goldman {\it et al.} {Phys. Rev. B} {\bf 79}, {024513} (2009)

\bibitem{Demuer}
   A. Demuer, A.T. Holmes, and D. Jaccard {J. Phys: Condens. Matter} {\bf 14}, {L529} (2002)
   
\bibitem{Saha}
	S. R. Saha {\it et al.}, {Phys. Rev. Lett.} {\bf 103}, 037005 (2009)


\bibitem{acknowledge}
 We would like to acknowledge helpful discussions with A. Kapitulnik, E. Gregoryanz, and A. Mackenzie.  T. Cuk would like to thank
 the Royal Holloway, University of London for the use of facilities
 and access to pressure cells. T. Cuk would also like to thank the Carnegie
 Insitution of Washington for the use of facilities, and partial support of this work.
 The Stanford work was supported by DOE Office of Science, Division of Materials Science, with
 contract DE-FG03-01ER45929-A001 and NSF grant DMR-0304981. Research at University of California, San Diego was supported by the National Nuclear Security Administration under the Stewardship Science Academic Alliance Program through the U. S. Department of Energy Grant No. DEFG52-06NA26205.

\end{thebibliography}
\end{document}